\documentclass[12pt]{article}
\usepackage{graphicx,wrapfig}
\usepackage{amsmath}
\usepackage{amsfonts}
\usepackage{amssymb}
\usepackage{epsfig}
\usepackage{subfigure}
\usepackage[square, comma, sort&compress]{natbib}

\begin{document}

\date{\empty}
\author{Y.~Frenkel$^{\mathrm{a}}$, I.~Gabitov$^{\mathrm{b}}$,
A.~Maimistov$^{\mathrm{c}}$, and V.~Roytburd$^{\mathrm{a}}$\\\\
$^{\mathrm{a}}$\textit{Department of Mathematical Sciences,
Rensselaer  Polytechnic}\\\textit{Institute, Troy, NY 12180-3590; }
$^{\mathrm{b}}$\textit{Department  of
Mathematics, }\\\textit{The University of Arizona, Tucson,  AZ 85721-0089};\\
$^{\mathrm{c}}$\textit{Department of Solid State Physics, Moscow Engineering
}\\\textit{Physics Institute, Kashirskoe sh. 31, Moscow, 115409 Russia}}
\title{Propagation of extremely short electromagnetic pulses in a doubly-resonant medium}
\maketitle
\begin{abstract}

Propagation of extremely short electromagnetic pulses in a homogeneous
doubly-resonant medium is considered in the framework of the total
Maxwell-Duffing-Lorentz model, where the Duffing oscillators (anharmonic oscillators
with cubic nonlinearities) represent the dielectric response of the
medium, and the Lorentz harmonic oscillators represent the magnetic response. The
wave propagation is governed by the one-dimensional Maxwell equations.

It is shown that the model
possesses a one-parameter family of traveling-wave solutions with the structure of single or multiple humps.
Solutions are parametrized by the velocity of propagation. The spectrum of possible velocities
is shown to be continuous on a small interval at the lower end of the spectrum; elsewhere
the velocities form a discrete set. A correlation between the number of humps and the velocity is
established. The traveling-wave solutions are found to be stable with respect to weak
perturbations. Numerical simulations demonstrate that the traveling-wave pulses collide in a
nearly elastic fashion.
\end{abstract}

\section{Introduction}

\noindent The recent demonstration of artificial materials (metamaterials)
with the left oriented triplet of electric $\vec{E}$, magnetic $\vec{H}$ and
wave vector $\vec{k}$ of electromagnetic
field~\cite{SSS01,Valent08,SCCYSDK05,ZPMOB05} stimulated studies of nonlinear
optical phenomena in such
materials~\cite{ASBZ04,ZSK03,LT05,AMSB05,SZK05,PS06,MG07}. Nonlinear dynamics
of extremely short optical pulses in left-handed materials was the subject of
particular interest in several recent
papers~\cite{LT05,SAMABCSB05,BVKR05,GILMSS06}. The first experimental
realization of the left-handed property based on the resonant response of the
artificial material to both electric and magnetic fields was described
in~\cite{SSS01}. To mention just one of the latest experimental achievements,
Valentine et al \cite{Valent08} were able to observe the negative refractive
index in the balk material in the \textit{optical} range. A theoretical
description of the electromagnetic wave interaction with such double resonance
materials (DRM) was considered in~\cite{ZH01,PHRS98,PHRS99,PSS02,MS02}.
Presence of two frequency intervals with different orientation of $(\vec
{E},\vec{H},\vec{k})$ triplets is a characteristic feature of such materials.

Most of the studies of electromagnetic pulse propagation in DRM has been
conducted in the slowly varying envelope approximation. On the other hand,
there is a broad area of nonlinear optical phenomena taking place in the limit
of extremely short pulses, when the slowly varying envelope approximation is
not valid~\cite{BK00}. The case of extremely short electromagnetic pulses
offers a new type of nonlinear interaction, when different frequency
components of electromagnetic pulses have different orientations of the
$(\vec{E},\vec{H},\vec{k})$ triplets.

The design of currently available DRM is based upon the use of embedded
metallic structures whose size is on the same order as the spatial size of an
extremely short electromagnetic pulse. Therefore a theoretical and numerical
investigation of the currently existing DRM would require 3D computer
simulation on Maxwell's equations that takes into account the strong
inhomogeneity of composite materials. Recently, there have been introduced
some qualitatively different approaches to design of DRM, including the use of
multilevel atoms~\cite{TM06,Krowne08a,Krowne08b}; the latter gives rise to a
spatially homogeneous medium. Possibilities of experimental realizations of
such an approach were recently discussed in~\cite{Yelin07,Yelin08}. As a first
step in the theoretical investigation of electrodynamics of homogeneous DRM in
this paper we study a simple model of a homogeneous doubly-resonant medium.
Even under such simplification, dynamics of extremely short pulses turn out to
be quite complex.

\section{Basic equations}

The system of equations that describe interaction of coherent light with a
medium consisting of molecules (considered as harmonic oscillators) is known
as the Maxwell-Lorentz model~\cite{AE75}. In this work we use a version of
the Maxwell-Lorentz system that is extended to account for simultaneous
magnetic and electric resonances, with the magnetic susceptibility being
linear, while the electric polarization being nonlinear. Consider the general
form Maxwell's equations:
\begin{align}
&  ~\nabla\times\vec{E}=-c^{-1}\vec{B}_{t},~~\;\nabla\times\vec{H}=-c^{-1}%
\vec{D}_{t}\\
&  ~\vec{B}=\vec{H}+4\pi\vec{M},~~\;\vec{D}=\vec{E}+4\pi\vec{P}
\nonumber\label{Maxwell:basic}%
\end{align}
For simplicity, we consider transverse electromagnetic plane waves propagating
along the $z$-axis with the electric field $\vec{E}=(E(z,t),0,0)$ and the
magnetic field $\vec{B}=(0,B(z,t),0).$ Then the Maxwell equations transform to
the scalar form:
\begin{align}
&  ~\frac{\partial E}{\partial z}+\frac{1}{c}\frac{\partial B}{\partial
t}=0,~~\frac{\partial H}{\partial z}+\frac{1}{c}\frac{\partial D}{\partial
t}=0\\
&  ~B=H+4\pi M,~~~\;D=E+4\pi P
\end{align}
which leads to
\begin{equation}
E_{z}+c^{-1}H_{t}=-4\pi c^{-1}M_{t},~~\;H_{z}+c^{-1}E_{t}=-4\pi c^{-1}P_{t}
\label{maxwell}%
\end{equation}
The system (\ref{maxwell}) must be closed by two additional equations
describing the interaction of the electric and magnetic fields with the DR
medium. As usual, it is convenient to avoid the  $4\pi$-factors by changing
the units for $M$ and $P$: $\tilde{M}=4\pi M,$ $\tilde{P}=4\pi P.$ In the
sequel we drop the tildes over $M$ and $P.$

Assume that the medium polarization is defined by the plasma oscillation of
electron density,
\begin{equation}
P_{tt}=\omega_{p}^{2}E\nonumber
\end{equation}
Here $\omega_{p}$ is an effective parameter characterizing polarizability of
the medium; in the case of metallic nanostructures it would be the effective
plasma frequency. To account for the dimensional quantization due to the
confinement of the plasma in nanostructures one should include the additional
term $\omega_{D}^{2}P$, where $\omega_{D}$ is the frequency of dimensional
quantization. We take into account nonlinearity in the lowest order of $P$,
which is $P^{3}$. A more accurate analysis, based on a quantum mechanical
approach~\cite{R97} and experimental measurements~\cite{DBNS04} confirms
validity of this assumption. Therefore we consider the modeling equation for
the medium polarization dynamics in the following form%
\begin{equation}
P_{tt}+\omega_{D}^{2}P+\kappa P^{3}=\omega_{p}^{2}E \label{polarization}%
\end{equation}
where $\kappa$ is a constant of anharmonisity. To account for magnetic
resonances we use the standard model~\cite{ZH01}
\begin{equation}
M_{tt}+\omega_{T}^{2}M=-\beta H_{tt} \label{magnetization}%
\end{equation}
Here $\beta$ is a constant characterizing magnetization.

We represent equations~(\ref{maxwell}), (\ref{polarization}) and
(\ref{magnetization}) in a dimensionless form by introducing $\tau=t/\tau_{0}$
($\tau_{0}=1/\omega_{p}$ is the characteristic time), $\eta=z/z_{0}$
($z_{0}=c\tau_{0}$ is the characteristic distance), $q=P/P_{0}$ ($P_{0}%
=\omega_{p}/\sqrt{\kappa}$ is the maximal achievable medium polarization). It
is convenient to normalize remaining variables as follows: $m=M/P_{0}$,
$e=E/P_{0}$, $h=H/P_{0}$. The system of dimensionless equations then takes the
following form:
\begin{align}
&  ~h_{\tau}+e_{\eta}=-m_{\tau},\nonumber\\
&  ~e_{\tau}+h_{\eta}=-q_{\tau},\nonumber\\
&  ~q_{\tau\tau}+\omega_{1}^{2}q+\gamma q^{3}=e\label{dimensionless:system}\\
&  ~m_{\tau\tau}+\omega_{2}^{2}m=-\beta h_{\tau\tau},\nonumber
\end{align}
where $\gamma=\kappa/\left(  \left|  \kappa\right|  \omega_{p}^{2}\right)  $,
$\omega_{1}=\omega_{D}/\omega_{p}$, $\omega_{2}=\omega_{T}/\omega_{p}$.

The system possesses the following conserved quantity:%
\begin{gather}
\frac{1}{2}\frac{\partial}{\partial\tau}\int\left[  \beta\omega_{2}^{2}\left(
e^{2}+\omega_{1}^{2}q^{2}+\frac{\gamma}{2}q^{4}\right)  +\beta\omega_{2}%
^{2}\left(  h+m\right)  ^{2}+\omega_{2}^{2}\left(  1-\beta\right)
m^{2}\right. \label{conserve}\\
\left.  +\beta\omega_{2}^{2}\left(  q_{\tau}\right)  ^{2}+\left[  m_{\tau
}+\beta h_{\tau}\right]  ^{2}\right]  d\eta=0\nonumber
\end{gather}
which is positive-definite for $\beta<1.$ For the traveling-wave solutions the
conservation relation (\ref{conserve}) yields conservation of electromagnetic
energy
\[
\frac{1}{2}\int\left(  e^{2}+h^{2}\right)  d\eta=\mathrm{const}%
\]
(see \cite{frenkel} $\ $for details). A natural question arises is whether the
system in (\ref{dimensionless:system}) possesses any solitary-wave solutions.
We address this issue in the following section.

\section{Solitary wave solutions}

Consider a traveling wave solution of (\ref{dimensionless:system}), i.e., a
solution that is a function of the variable $\zeta=\tau-\eta/V.$ Then the PDEs
in (\ref{dimensionless:system}) become ODEs, and one obtains the following
system:
\begin{align}
h^{\prime}-e^{\prime}/V  &  =-m^{\prime}\label{ODE:basic:system1}\\
e^{\prime}-h^{\prime}/V  &  =-q^{\prime}\label{ODE:basic:system2}\\
q^{\prime\prime}+\omega_{1}^{2}q+\gamma q^{3}  &  =e \label{ODE:basic:system3}%
\\
m^{\prime\prime}+\omega_{2}^{2}m  &  =-\beta h^{\prime\prime}
\label{ODE:basic:system4}%
\end{align}
Upon the integration of equations~(\ref{ODE:basic:system1})
and~(\ref{ODE:basic:system2}) once, we get the algebraic conservation
relations
\begin{align*}
Vh-e  &  =-mV+R\\
-h+eV  &  =-qV+S
\end{align*}
We are interested in a traveling-wave solution \textit{on the zero background},
hence $h=m=q=e=0$ at $\pm\infty;$ therefore the constants of integration
$R=S=0.$ This yields the following expressions for $h$ and $e$
\begin{align}
h  &  =a_{1}m+a_{2}q\label{ODE:reduced:system1}\\
e  &  =a_{2}m+a_{1}q \label{ODE:reduced:system2}%
\end{align}
where
\begin{align}
a_{1}=V^{2}\left(  1-V^{2}\right)  ^{-1},\quad a_{2}=V\left(  1-V^{2}\right)
^{-1} \label{coefficients:a1:a2}%
\end{align}

We insert expressions~(\ref{ODE:reduced:system1})
and~(\ref{ODE:reduced:system2}) for $h$ and $e$ into the
equations~(\ref{ODE:basic:system3}) and~(\ref{ODE:basic:system4}) for $q$ and
$m$ and obtain the following system of second order equations:
\begin{align*}
q^{\prime\prime}+\left(  \omega_{1}^{2}-a_{1}\right)  q-a_{2}m+\gamma q^{3}
&  =0\\
\beta a_{2}q^{\prime\prime}+\left(  1+\beta a_{1}\right)  m^{\prime\prime
}+\omega_{2}^{2}m  &  =0
\end{align*}

This system can be diagonalized with respect to the second derivatives
\begin{align}
Q^{\prime\prime}+A_{11}Q+A_{12}M+\gamma Q^{3}  &  =0 \label{diagonalized:form}%
\\
M^{\prime\prime}+A_{21}Q+A_{22}M  &  =0\nonumber
\end{align}
by the means of the transformation
\[
\left[
\begin{array}
[c]{l}%
q\\
m
\end{array}
\right]  =\left[
\begin{array}
[c]{ll}%
1 & 0\\
\dfrac{-\beta a_{2}}{1+\beta a_{1}} & \dfrac{\omega_{2}\sqrt{\beta}}{1+\beta
a_{1}}%
\end{array}
\right]  \left[
\begin{array}
[c]{l}%
Q\\
M
\end{array}
\right]
\]
The matrix $A$ in (\ref{diagonalized:form}) is symmetric $A_{12}=A_{21}:$
\begin{equation}
A=\left[
\begin{array}
[c]{ll}%
\omega_{1}^{2}-a_{1}+\dfrac{\beta a_{2}^{2}}{1+\beta a_{1}} & -\dfrac
{a_{2}\omega_{2}\sqrt{\beta}}{1+\beta a_{1}}\\
-\dfrac{a_{2}\omega_{2}\sqrt{\beta}}{1+\beta a_{1}} & \dfrac{\omega_{2}^{2}%
}{1+\beta a_{1}}%
\end{array}
\right]  \label{matr_A}%
\end{equation}

Instead of the second order system (\ref{diagonalized:form}) we will consider
the following equivalent $4\times4$ first order system
\begin{equation}
\frac{d}{d\zeta}\left[
\begin{array}
[c]{c}%
Q\\
M\\
Q_{1}\\
M_{1}%
\end{array}
\right]  =\left[
\begin{array}
[c]{cccc}%
0 & 0 & 1 & 0\\
0 & 0 & 0 & 1\\
-A_{11} & -A_{12} & 0 & 0\\
-A_{21} & -A_{22} & 0 & 0
\end{array}
\right]  \left[
\begin{array}
[c]{c}%
Q\\
M\\
Q_{1}\\
M_{1}%
\end{array}
\right]  -\left[
\begin{array}
[c]{c}%
0\\
0\\
\gamma Q^{3}\\
0
\end{array}
\right]  \label{first}%
\end{equation}
Obviously $[0,0,0,0]$ (the zero background) is the only equilibrium solution
(the critical point) of the system. The pulse solutions are the trajectories
of the system (\ref{first}) that start and end at the equilibrium (homoclinic
orbits). Thus, the investigation of solitary pulses is mathematically
equivalent to studying homoclinic solutions.

\begin{figure}
\begin{minipage}[b]{0.5\linewidth} 
\centering
\includegraphics[width=7.6cm]{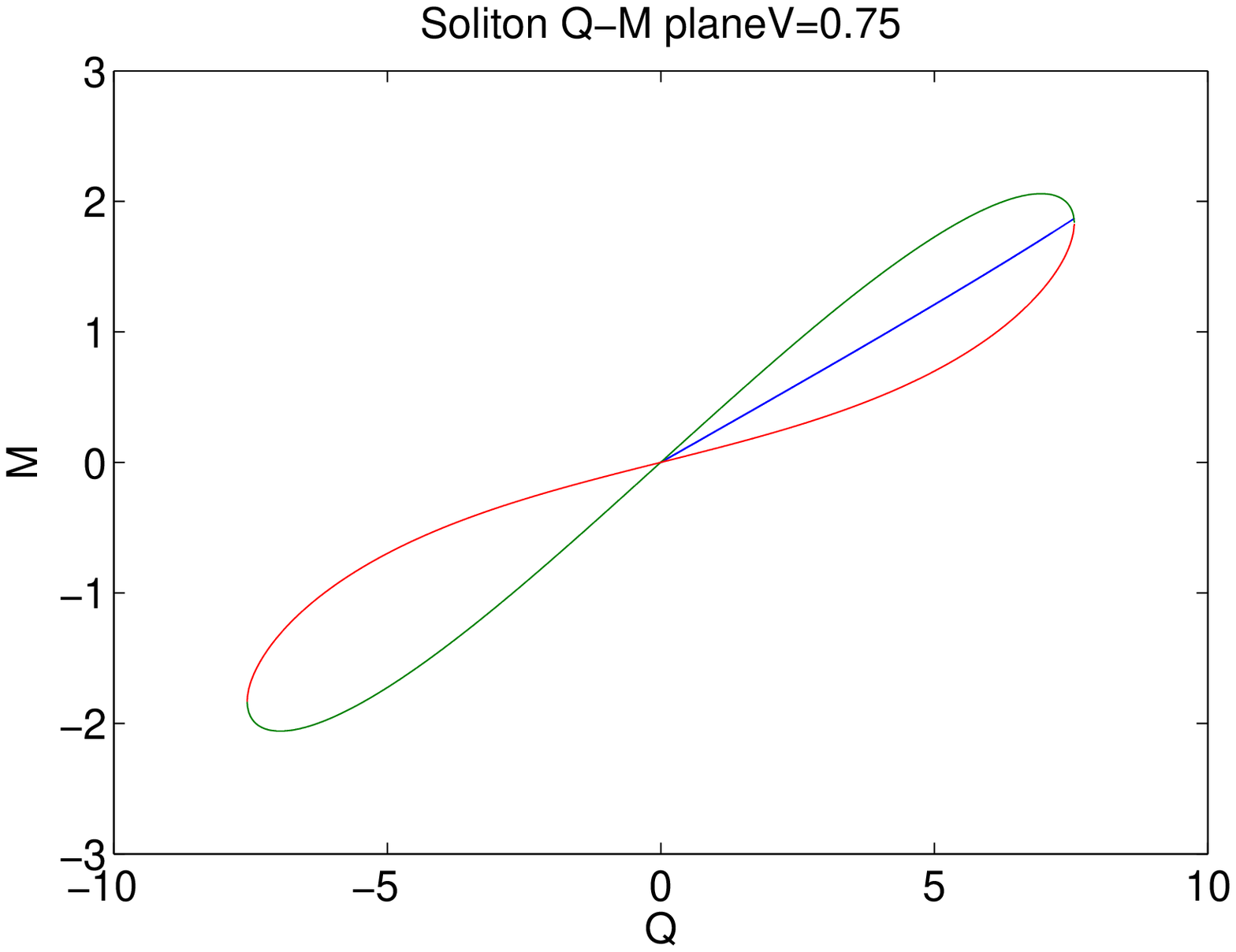}
\end{minipage}
\begin{minipage}[b]{0.5\linewidth}
\centering
\includegraphics[width=7.4cm]{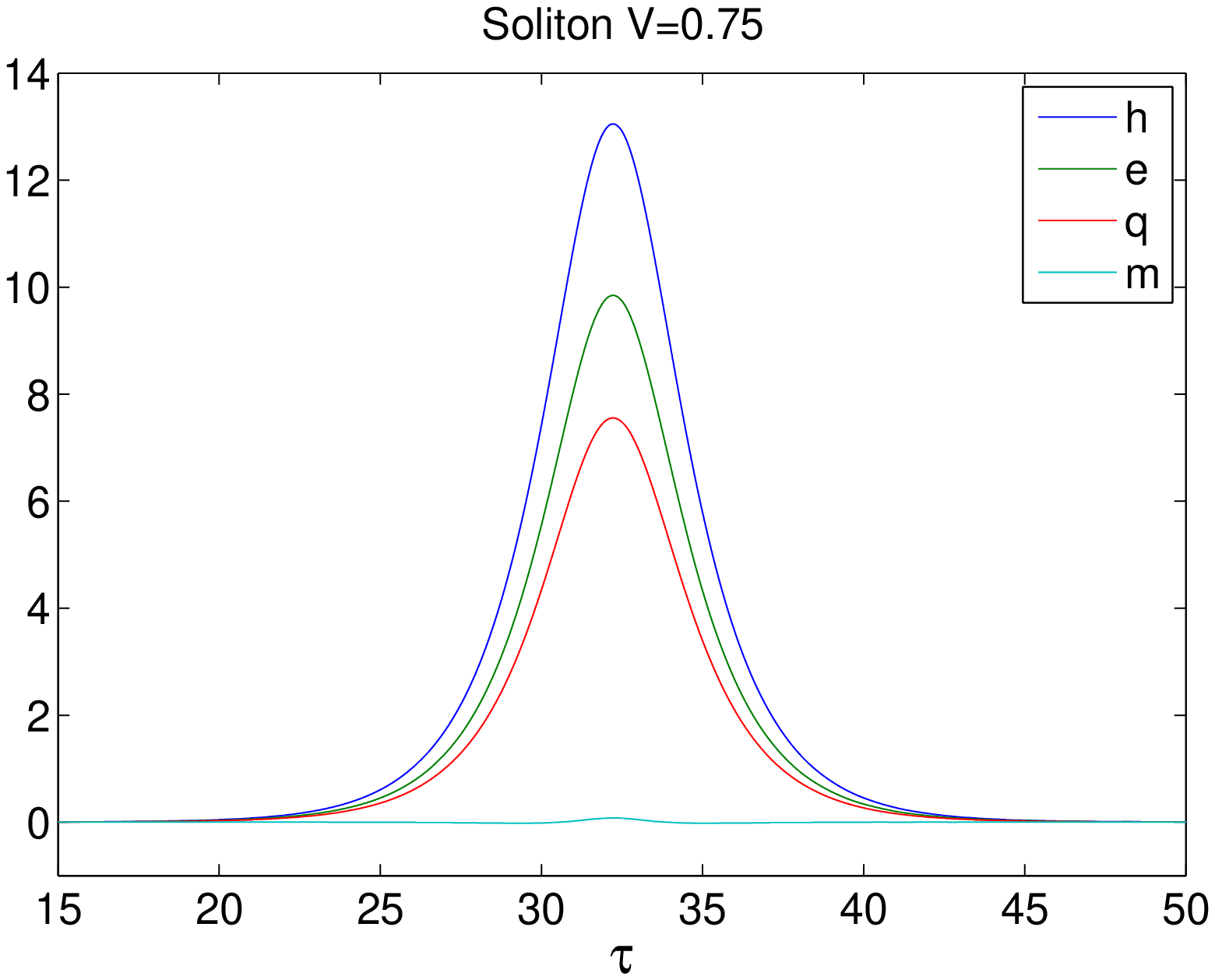}
\end{minipage}
\caption{The left figure shows the
$E=0$ cross-section of the potential energy landscape $U(Q,M)=0$. The
Newtonian particle trajectory corresponds to a one-hump solution presented
in the right figure.}
\label{Fig2}
\end{figure}


\section{Structure of solitary waves}

To investigate the structure of homoclinic solutions, we linearize the system
in (\ref{first}) near the critical point $Q=M=Q_{1}=M_{1}=0$:
\begin{equation}
\frac{d}{d\zeta}\left[
\begin{array}
[c]{c}%
Q\\
M\\
Q_{1}\\
M_{1}%
\end{array}
\right]  =\left[
\begin{array}
[c]{cccc}%
0 & 0 & 1 & 0\\
0 & 0 & 0 & 1\\
-A_{11} & -A_{12} & 0 & 0\\
-A_{21} & -A_{22} & 0 & 0
\end{array}
\right]  \left[
\begin{array}
[c]{c}%
Q\\
M\\
Q_{1}\\
M_{1}%
\end{array}
\right]  :=\tilde{A}\left[
\begin{array}
[c]{c}%
Q\\
M\\
Q_{1}\\
M_{1}%
\end{array}
\right]  \label{a-tilde}%
\end{equation}
The characteristic equation of the matrix $\tilde{A}$ on the right-hand side
is given by
\[
p^{4}+\left(  A_{11}+A_{22}\right)  p^{2}+A_{11}A_{22}-A_{21}A_{12}=0
\]
Therefore, the values of $p^{2}$ coincide with the eigenvalues of the matrix
$-A.$ It is easy to see that%
\begin{equation}
\det A=\dfrac{\left(  \omega_{1}^{2}-a_{1}\right)  \omega_{2}^{2}}{1+\beta
a_{1}} \label{det}%
\end{equation}
Thus, the condition
\begin{equation}
\omega_{1}^{2}-a_{1}<0 \label{necessary}%
\end{equation}
makes $\det A<0,$ causing $A$ to have eigenvalues of opposite signs, which is
a necessary condition for the existence of homoclinic orbits. Indeed, in the
case of $A$ having eigenvalues of the opposite signs, the $4\times4$ matrix
$\tilde{A}$ has two pure imaginary eigenvalues (square roots of the negative
eigenvalue of $-A$) and one negative, and one positive eigenvalues. Therefore
the nonlinear system has one-dimensional stable and unstable manifolds, and a
two-dimensional center manifold (corresponding to the imaginary eigenvalues).

\begin{figure}
\begin{minipage}[b]{0.5\linewidth} 
\centering
\includegraphics[width=7.5cm]{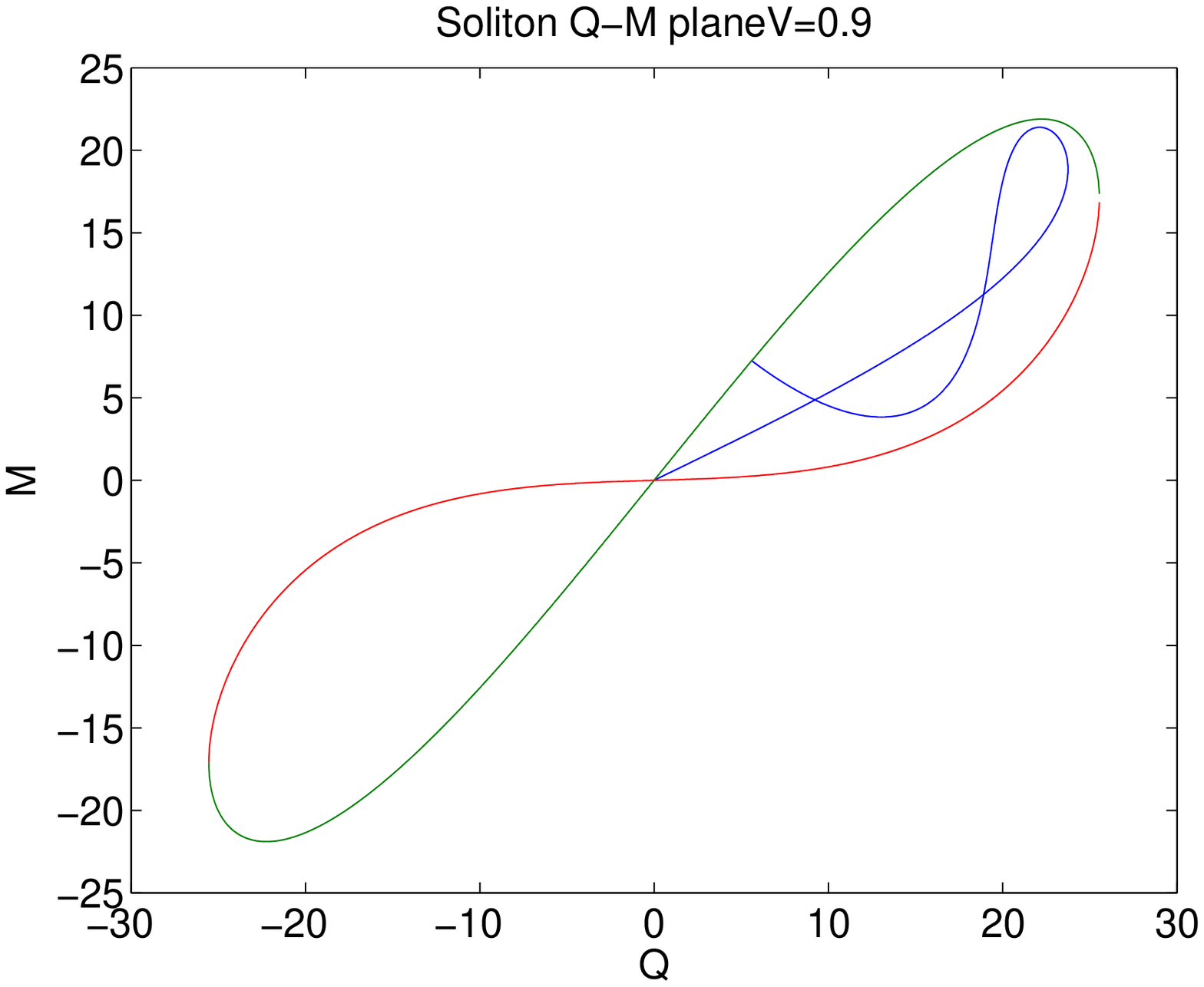}
\end{minipage}
\begin{minipage}[b]{0.5\linewidth}
\centering
\includegraphics[width=7.5cm]{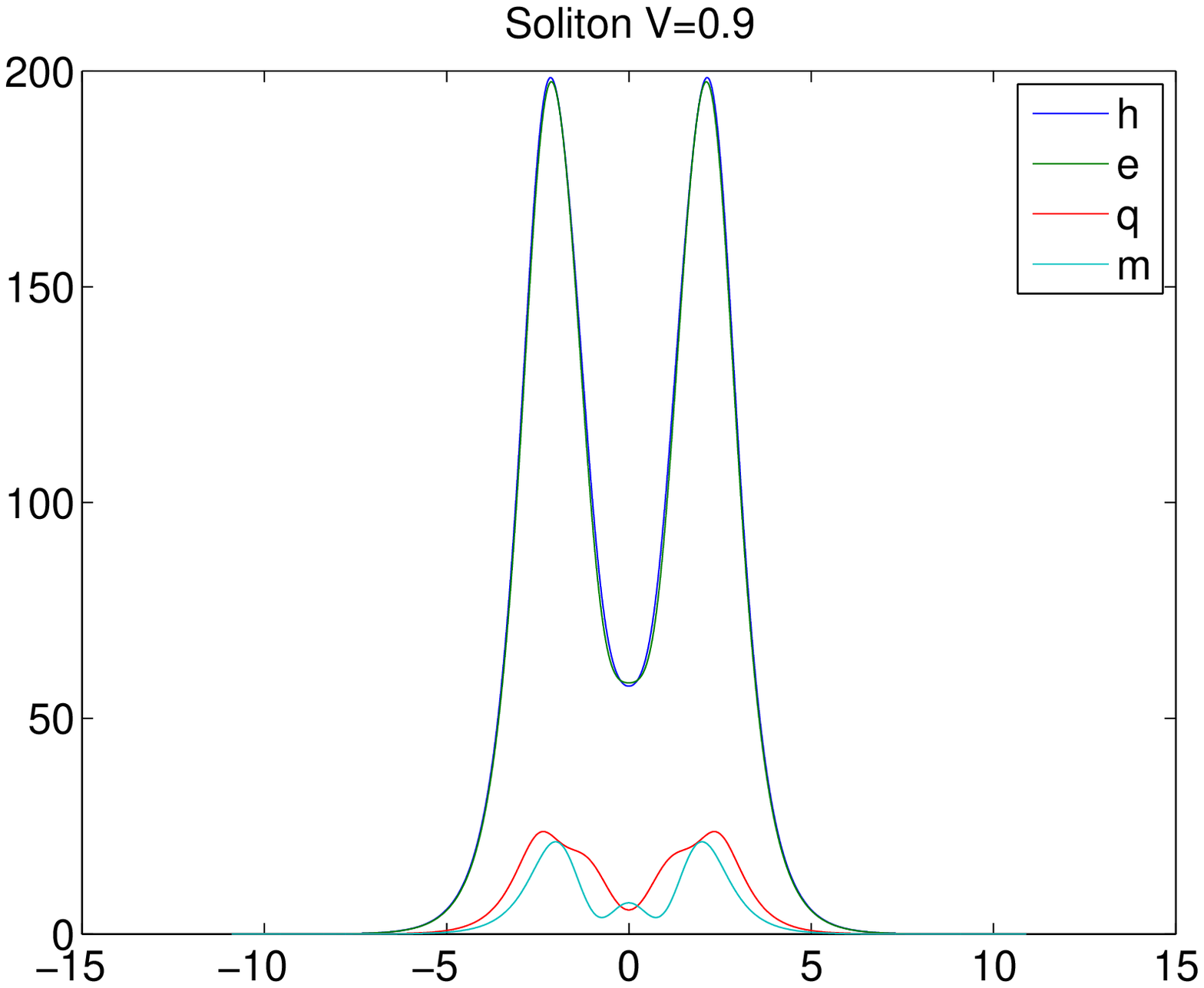}
\end{minipage}
\caption{The Newtonian particle trajectory (left) corresponds to the two-hump solitary wave (right).}
\label{Fig3}
\end{figure}

It was first noticed in \cite{GILMSS06} that the nonlinear system in
(\ref{first}) has Hamiltonian structure. If the kinetic, $E,$ and potential,
$U,$ energies and the Hamiltonian, $H,$ are introduce as follows
\begin{align}
E  &  =\frac{1}{2}\left(  Q_{1}^{2}+M_{1}^{2}\right)  , \label{Kinetic:energy}%
\\
U  &  =A_{12}QM+\frac{1}{2}\left(  A_{11}Q^{2}+A_{22}M^{2}\right)
+\frac{\gamma}{4}Q^{4},\label{Potential:energy}\\
H  &  =E+U \label{Hamiltonian}%
\end{align}
then the system (\ref{first}) takes the form
\begin{align*}
\partial_{\zeta}Q_{1}  &  =-\partial H/\partial Q,\quad\partial_{\zeta}%
M_{1}=-\partial H/\partial M,\\
\partial_{\zeta}Q  &  =\partial H/\partial Q_{1},\quad\partial_{\zeta
}M=\partial H/\partial M_{1}.
\end{align*}

Since the Hamiltonian is a conserved quantity, $\partial_{\zeta}H=0,$ any
trajectory issued from the critical point $\left[  0,0,0,0\right]  $ stays on
the zero energy level surface $H=0$ for all time$.$ Note that the surface
$H=0$ is a 3D manifold in $\mathbb{R}^{4}.$ The intersection of this 3D
hypersurface with the hyperplanes $Q_{1}=0$ and $M_{1}=0$ is a curve $\Gamma$
in the $QM$-plane
\begin{equation}
U(Q,M)=A_{12}QM+\frac{1}{2}\left(  A_{11}Q^{2}+A_{22}M^{2}\right)
+\frac{\gamma}{4}Q^{4}=0 \label{Gamma}%
\end{equation}
(the figure-eight shaped curve on the left in Fig.~\ref{Fig2} and \ref{Fig3}). If for a given
$V$ there exits a homoclinic trajectory of (\ref{first}), then on this
trajectory $E+U=0$ and since $E\geq0$, necessarily $U\leq0$. At the extrema of
$U$ its gradient is zero:
\[
\frac{\partial U}{\partial M}=A_{22}M+A_{12}Q=0,\quad\frac{\partial
U}{\partial Q}=A_{12}M+A_{11}Q+\gamma Q^{3}=0
\]
By eliminating $M$ from the equations above, we obtain the cubic equation
\[
-\frac{A_{12}^{2}}{A_{22}}Q+A_{11}Q+\gamma Q^{3}=0
\]
whose roots are easily found:
\[
Q=0,\quad Q=\pm\sqrt{\frac{-\det A}{A_{22}}}=\pm\sqrt{a_{1}-\omega_{1}^{2}}.
\]
Thus, $\nabla U=0$ at the points
\[
\left(  0,~0\right)  ,~~\left(  \pm\sqrt{a_{1}-\omega_{1}^{2}},~\pm\frac
{a_{2}}{\omega_{2}}\sqrt{\beta\left(  a_{1}-\omega_{1}^{2}\right)  }\right)
\]
which are real if
\[
a_{1}=\frac{V^{2}}{1-V^{2}}>\omega_{1}^{2}%
\]
thus producing the figure eight level curves. We already encountered this
inequality above, see (\ref{necessary}). After some algebra it can be
rewritten as the following constraints on the traveling wave velocity:
\begin{equation}
V_{0}<V<1,\quad V_{0}=\sqrt{\omega_{1}^{2}/(1+\omega_{1}^{2})}
\label{V-bounds}%
\end{equation}
Thus, a possible velocity of the propagating pulse is bounded below.

\section{Numerical study of solitary waves\label{num_waves}}

The nonlinear system (\ref{first}) has the time reversal symmetry; therefore,
if $[Q,M,Q_{1},$ $M_{1}](t):=\mathbf{u}(t)$ is a homoclinic orbit,
$[Q,M,-Q_{1},-M_{1}](-t)$ also is (recall that $Q_{1}$ and $M_{1}$ are time
derivatives of $Q$ and $M$).

\begin{figure}[ht]
\centering
\subfigure[Number of
solitons per bin: 100 bins is depicted.]{
\includegraphics[width=0.44\textwidth]{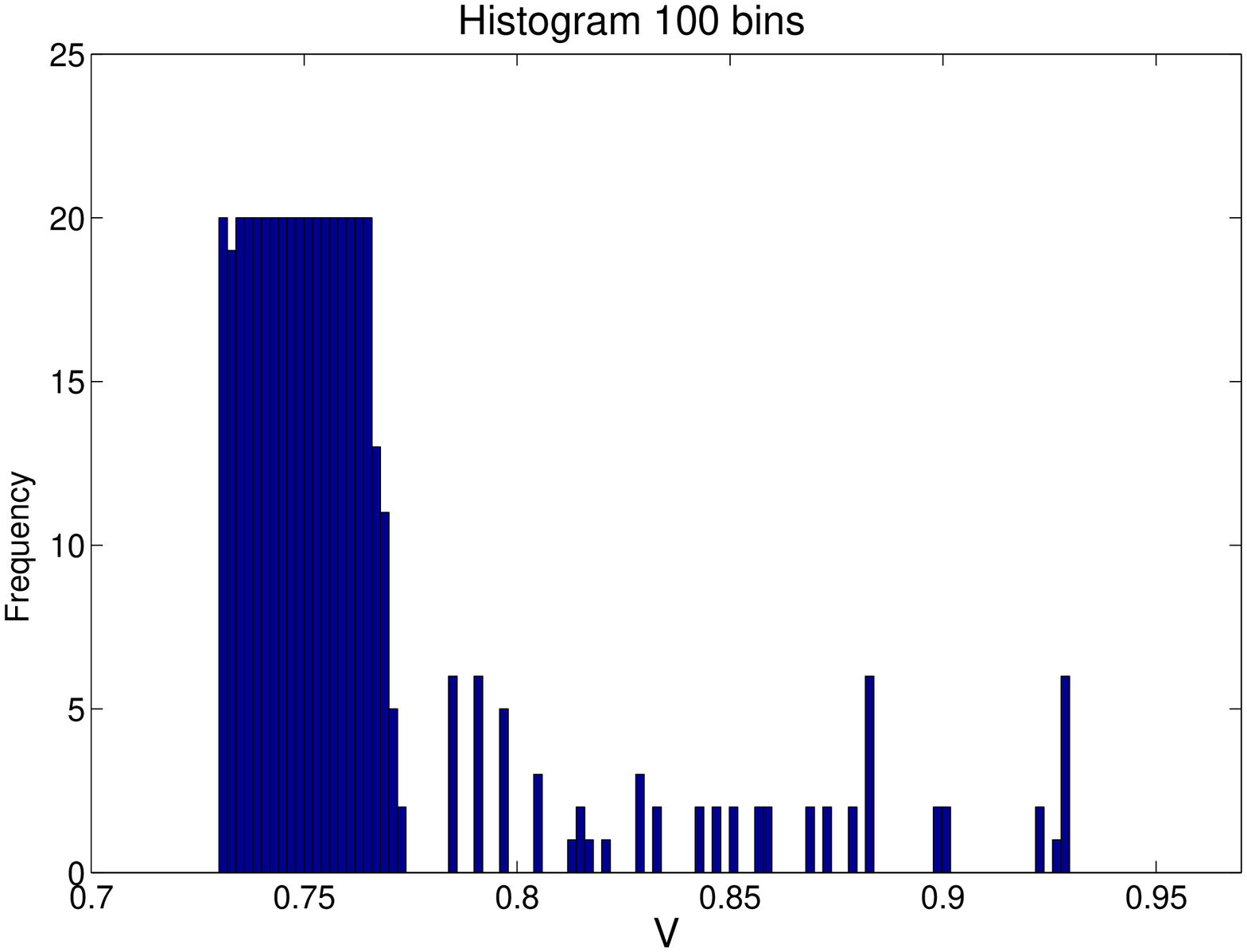}
\label{Fig4}
}\hspace{0.5cm}
\subfigure[The hump distribution: number of humps vs. the velocity. ]{
\includegraphics[width=0.47\textwidth]{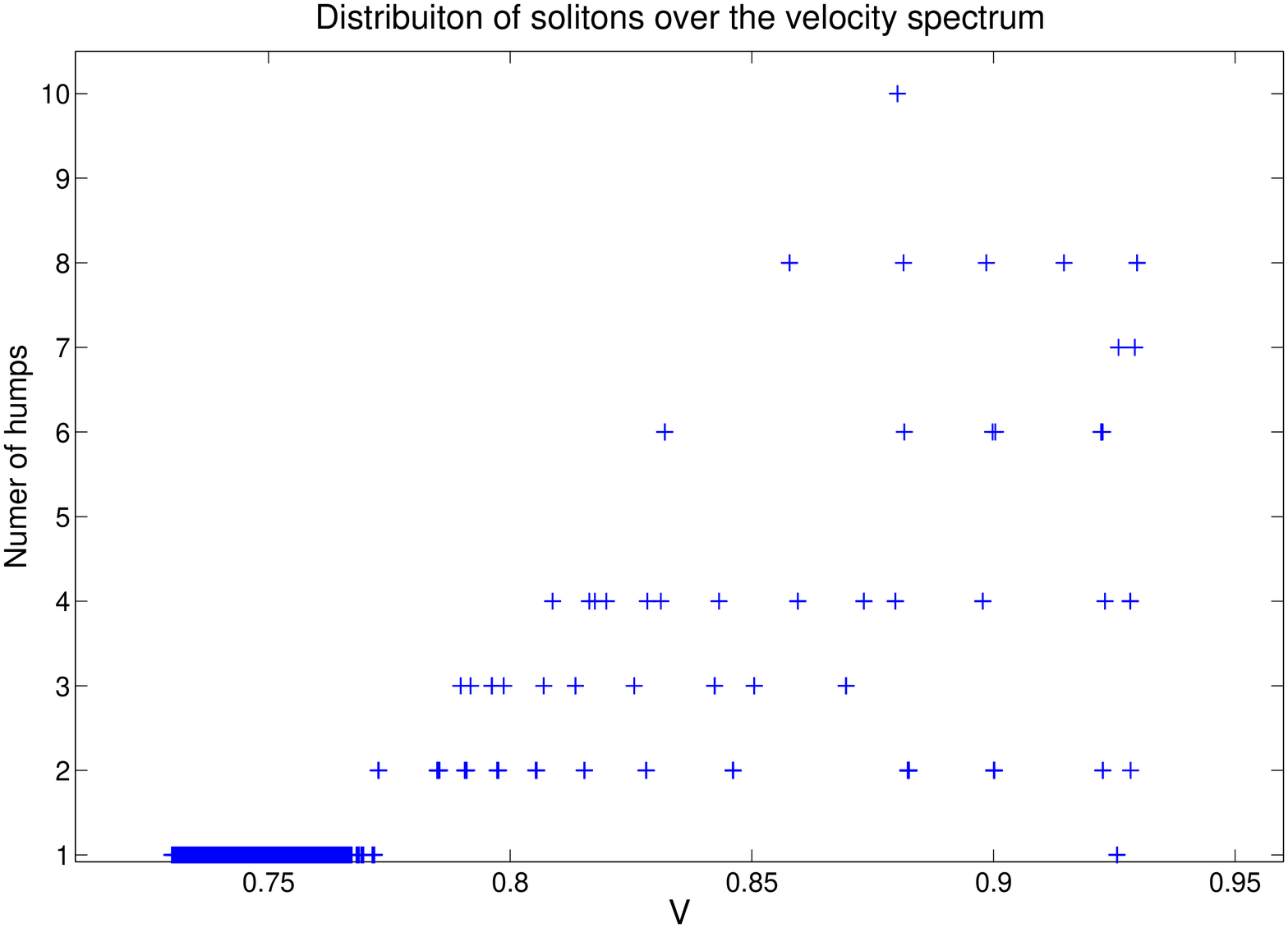}
\label{thehumpdistribution}
}
\caption[]{Statistics of solitary wave solutions.}
\end{figure}

A priori it is not clear why any homoclinic
solution would possess this symmetry, and it is quite likely that there exist
non-symmetric homoclinic orbits; we plan to investigate them elsewhere.
The characteristic property of a time reversal orbit is that at the symmetry point
$Q_{1}=M_{1}=0,$ and consequently the kinetic energy $E$ must be zero; i.e.,
the symmetry point lies on the curve $\Gamma,$ see (\ref{Gamma}). Moreover at
the symmetry point the trajectory is orthogonal to $\Gamma$ (for an
illustration, see the $QM$ diagrams on the left of Fig.~\ref{Fig2} and
Fig.~\ref{Fig3}).
\begin{figure}[tbhtbhtbh]
\centering
\includegraphics[width=0.49\textwidth]{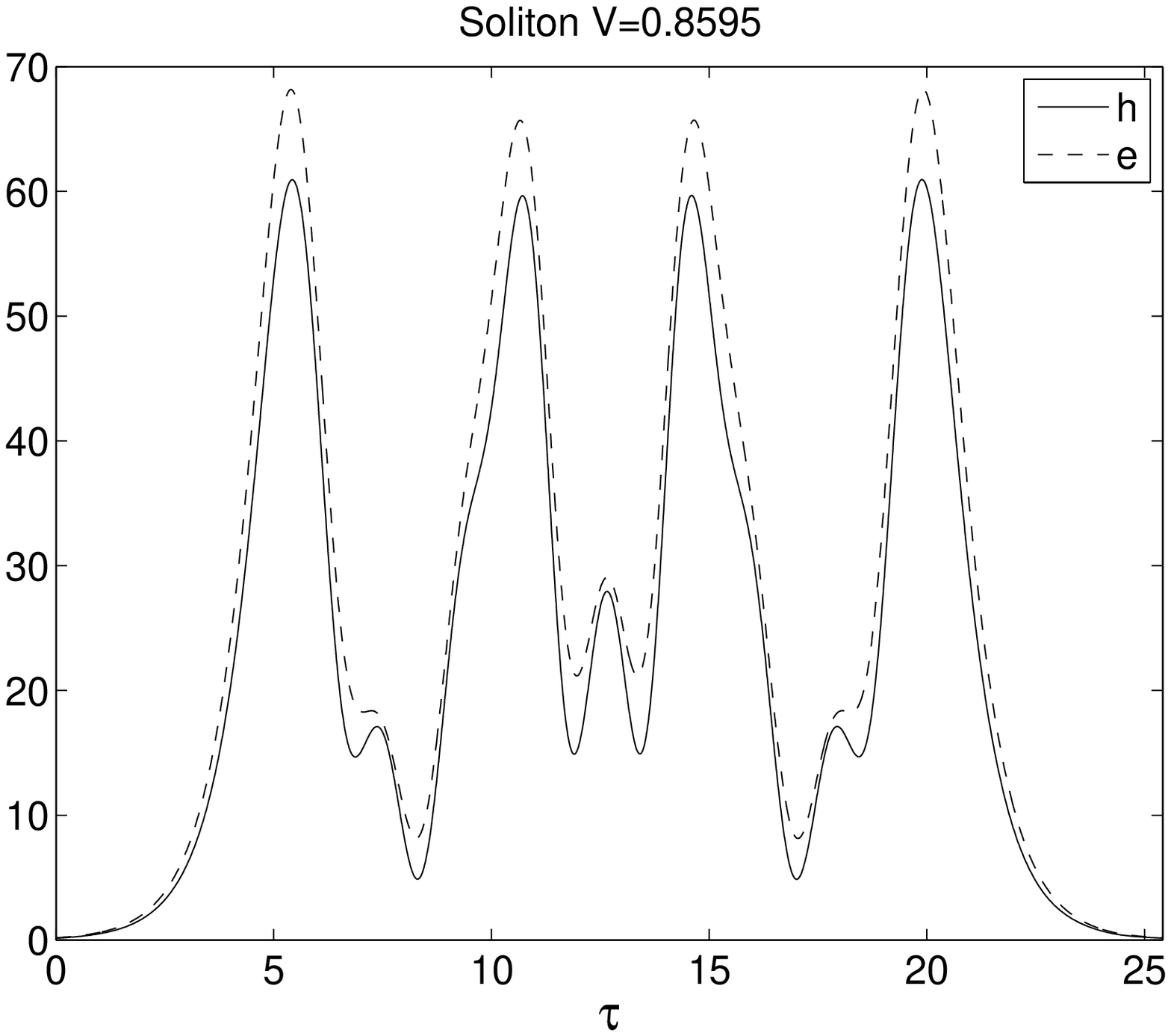}
\includegraphics
[width=0.49\textwidth]{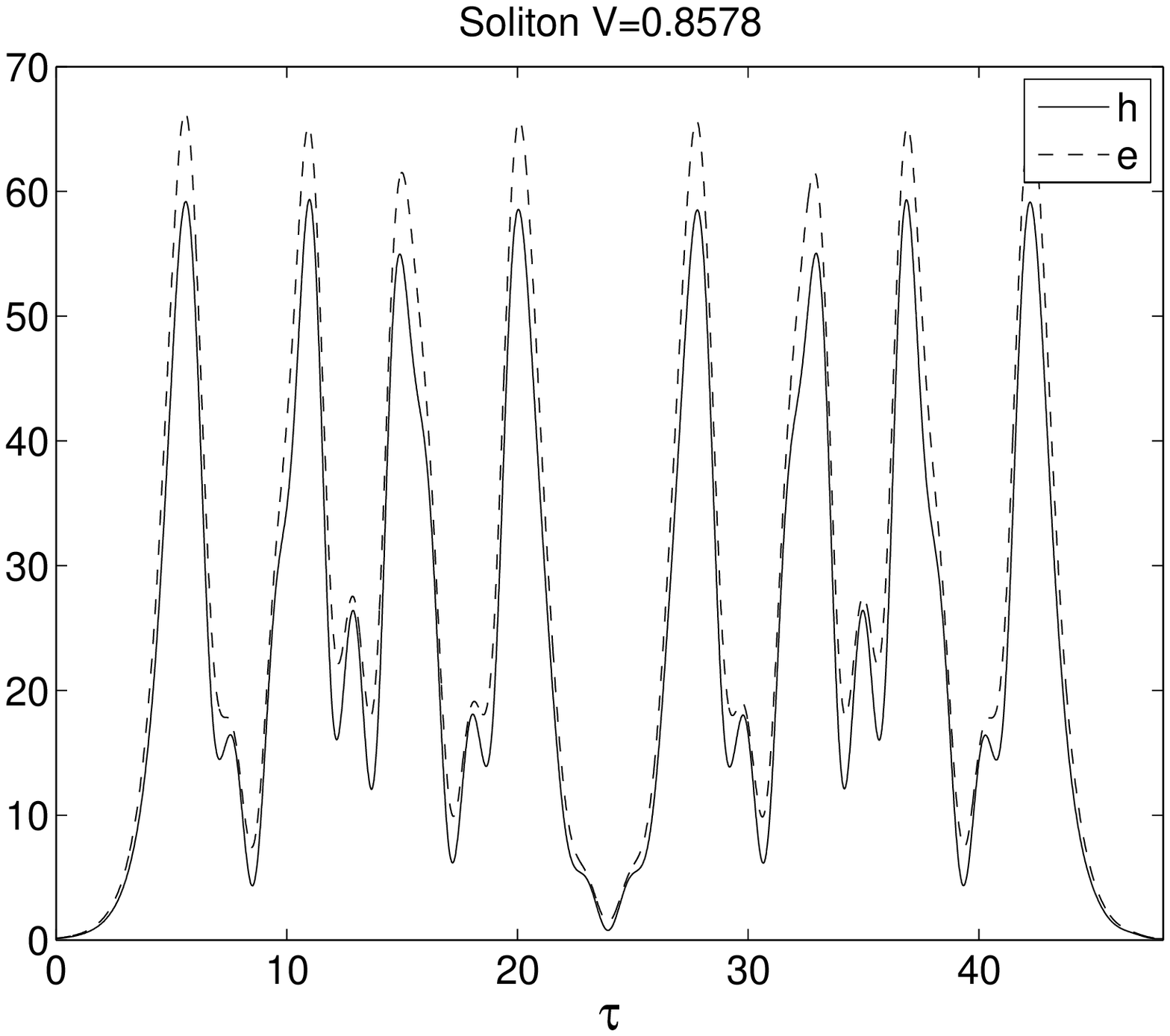}\caption{Solitary waves examples
$V=0.84322$ and $V=0.91461$. The left and right figures illustrate four-hump
and eight-hump solitary wave solutions. }
\label{profiles:1-4}
\end{figure}

Our search algorithm for finding solitary-wave solutions is based on the
following minimization idea. If for a given value of the propagation velocity
$V$ there exists a homoclinic orbit with the time-reversal symmetry, then at
some point both the kinetic and potential energies are zero. The algorithm
takes the initial condition $\mathbf{u}_{0}$ from a domain $S$ on the
zero-energy surface, near the critical point $(0,0,0,0)$ and in the direction
close to that of the unstable eigenvector of the linearized problem. Then the
following optimization problem is posed: Determine
\begin{equation}
\Phi(V)=\min_{\mathbf{u}_{0}\in S}\min_{\zeta_{0}<\zeta<\zeta_{0}+\tau_{0}
}E[\mathbf{u}(\zeta|\mathbf{u}_{0})] \label{optimization}
\end{equation}
\begin{wrapfigure}{r}{0.5\textwidth}
\begin{center}
\includegraphics[width=0.45\textwidth]{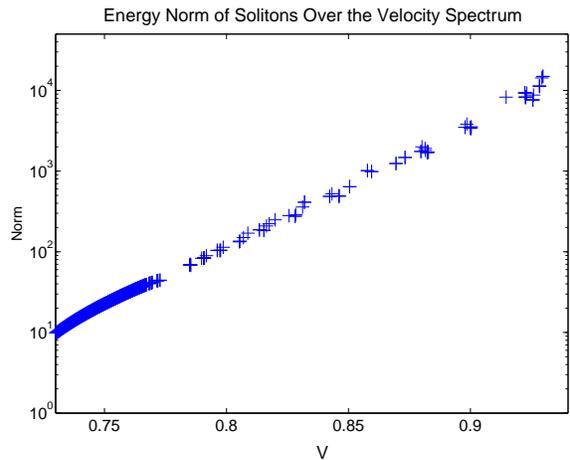}
\end{center}
\caption{The energy per hump vs. the velocity}
\label{Energy:per:single:hump}
\end{wrapfigure}
where $E$ is the kinetic energy, $\mathbf{u}(\zeta|\mathbf{u}_{0})$ is the
solution of (\ref{first}) with the initial condition $\mathbf{u}%
(0|\mathbf{u}_{0})=\mathbf{u}_{0},$ recall that $\mathbf{u}=[Q,M,Q_{1}%
,M_{1}].$ The parameter $\tau_{0}$ is the expected ''width'' of the pulse.
Since $E=0$ at $\zeta=0$ we take $\zeta>\zeta_{0}$ to obtain a nontrivial
solution for the energy minimization problem. Computation of any particular
value of $\min_{\zeta_{0}<\zeta<\zeta_{0}+\tau_{0}}E[\mathbf{u}(\zeta
|\mathbf{u}_{0})]$ involves a numerical solution of the nonlinear system of
ODEs.

\begin{figure}[ht]
\begin{center}
\includegraphics[width=0.6\textwidth]{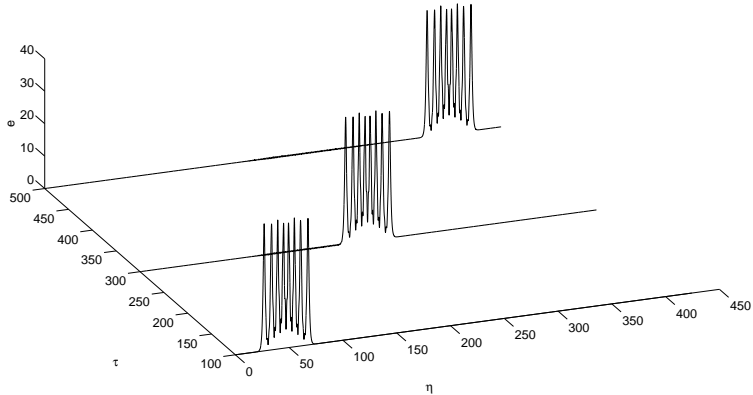}
\end{center}
\caption[]{Stable propagation of a
eight-hump solitary wave; $V=0.822$.}
\label{fig:EV6prop}
\end{figure}

The search of the optimal initial datum is stochastic and is organized
via a version of simulated annealing \cite{simannealing}. On each step the
initial datum is obtained by sampling a random distribution with the density
determined by the results of the previous step (see \cite{frenkel,yfvr09} for
more detail). If $\Phi(V)=0$ then there exists a homoclinic solutions with
velocity $V.$

When the kinetic energy possesses several local minima along the trajectory
the corresponding solitary wave has the multi-hump structure.
Figures~\ref{Fig2} and \ref{Fig3} illustrate this phenomenon. The figure-eight
shaped curves on the left correspond to the $E=0$ cross-section of the
potential energy landscapes; the curves inside the domains represent the
Newtonian particle trajectories in the $QM$ configuration space. The graphs on
the right show the profiles of the corresponding solitary wave solutions.
Fig.~\ref{Fig2} illustrates a typical one-hump solution. In contrast, the
trajectory shown in Fig.~\ref{Fig3} has a point of the nearest approach to the
boundary where the kinetic energy attains a local minimum. The resulting
solution has a two-hump structure. Multi-hump solutions correspond to more
complicated trajectories. Each of these trajectories has the return point at
which it has the normal incidence with the $E=0$ contour.

\begin{figure}[tbhtbh]
\includegraphics[width=0.49\textwidth]{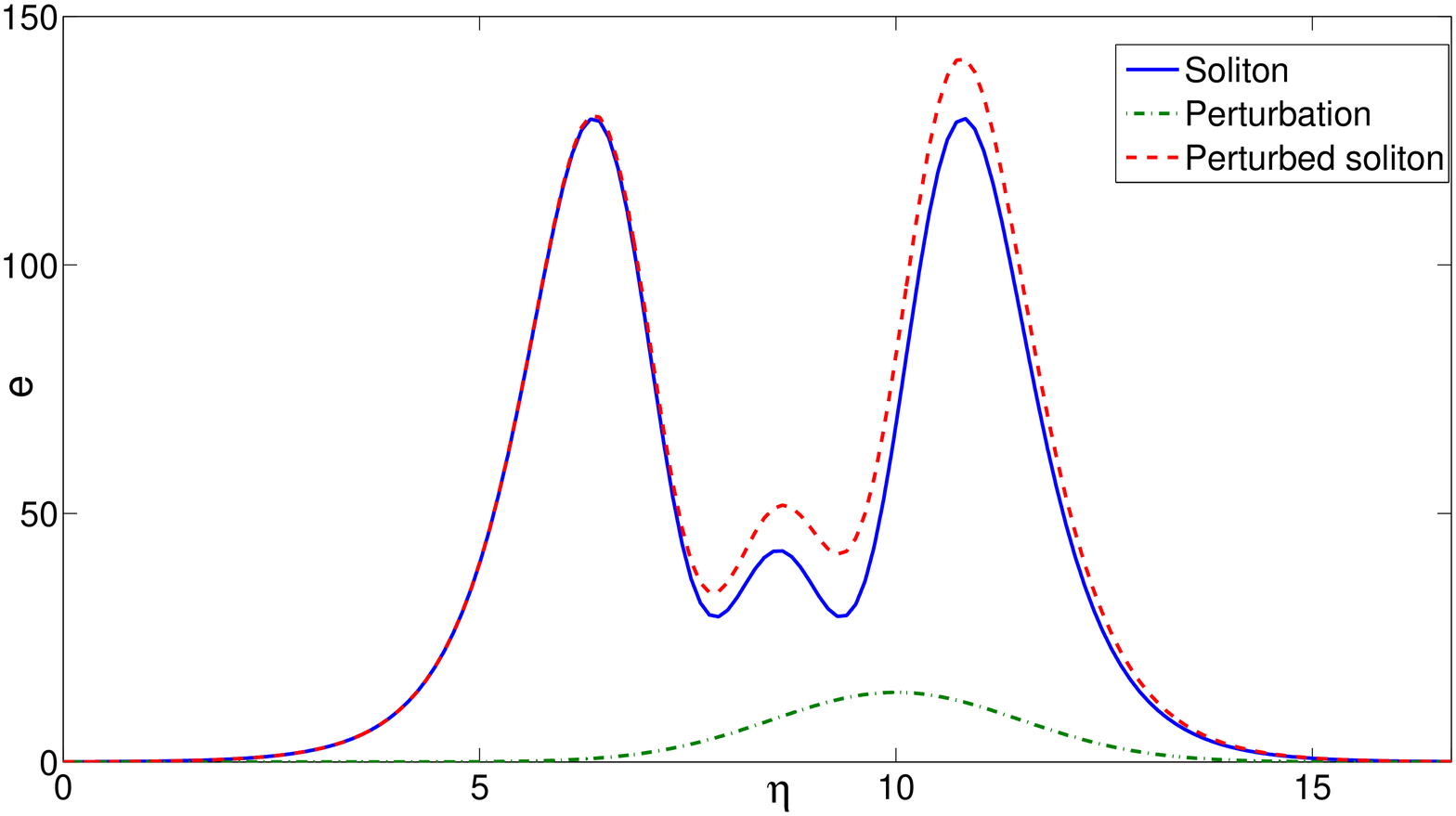}
\includegraphics
[width=0.49\textwidth]{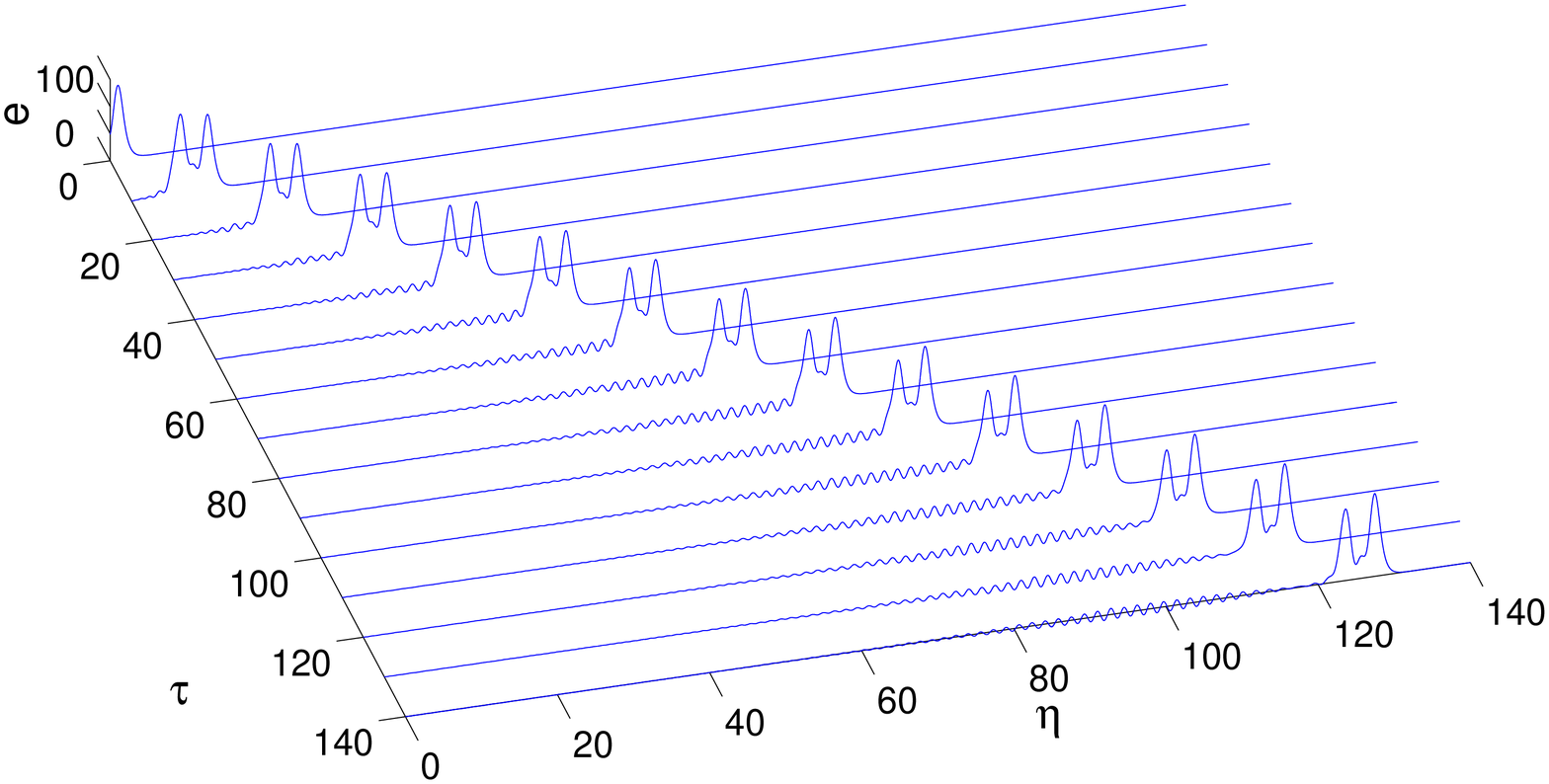}\caption{The initial solitary wave
pulse with a perturbation added (left); Evolution of this pulse governed by
the PDEs (right).}%
\label{Fig6}%
\end{figure}

For the fixed set of physical parameter values, the shape of the potential
energy landscape is controlled by the pulse velocity $V$ via the coefficients
$a_{1}$ and $a_{2}$~in~(\ref{coefficients:a1:a2}). We investigated numerically
the set $\frak{V}$ of values of $V$ which give rise to homoclinic orbits; in
some sense one might think of these $V$s as the ''spectrum'' of the problem.
For numerous applications with soliton-like solutions the velocity value is
known to change continually (a continuous spectrum). However, for the
Maxwell-Duffing model under consideration our numerical investigation
demonstrates that the spectrum $\frak{V}$ contains both an interval of a
continuous spectrum and a discrete subset of parameter values $V$ for which a
wave solution exists. One of the principal issues is to understand the
correspondence between types of solitary wave solutions and values of
$V\in\frak{V}$.

We first investigated numerically the distribution density of the values of
$V$, which give rise to homoclinic orbits. For all numerical computations of
this section we adopted the following values of the nondimensional physical
parameters:
\begin{equation}
\omega_{1}=1,\;\omega_{2}=5,\;\gamma=0.01,\;\beta=0.5 \label{params}%
\end{equation}
For $\omega_{1}=1$ the allowable range of values of $V$ from (\ref{V-bounds})
is given by $1/\sqrt{2}<V<1.$ The plot in Fig.~\ref{Fig4} illustrates the
density distribution of $V\in\frak{V}$ on the interval $[0.73,0.95]$. The
search algorithm tested potential values of $V$ on the grid $\delta
V=10^{-4}.$ The plot depicts the number of ``successful'' homoclinic orbits
per velocity interval $\Delta V$ (a ``bin''); in this particular case the
value has been chosen as $\Delta V=0.002$.

\begin{figure}[tbhtbh]
\includegraphics[width=0.49\textwidth]{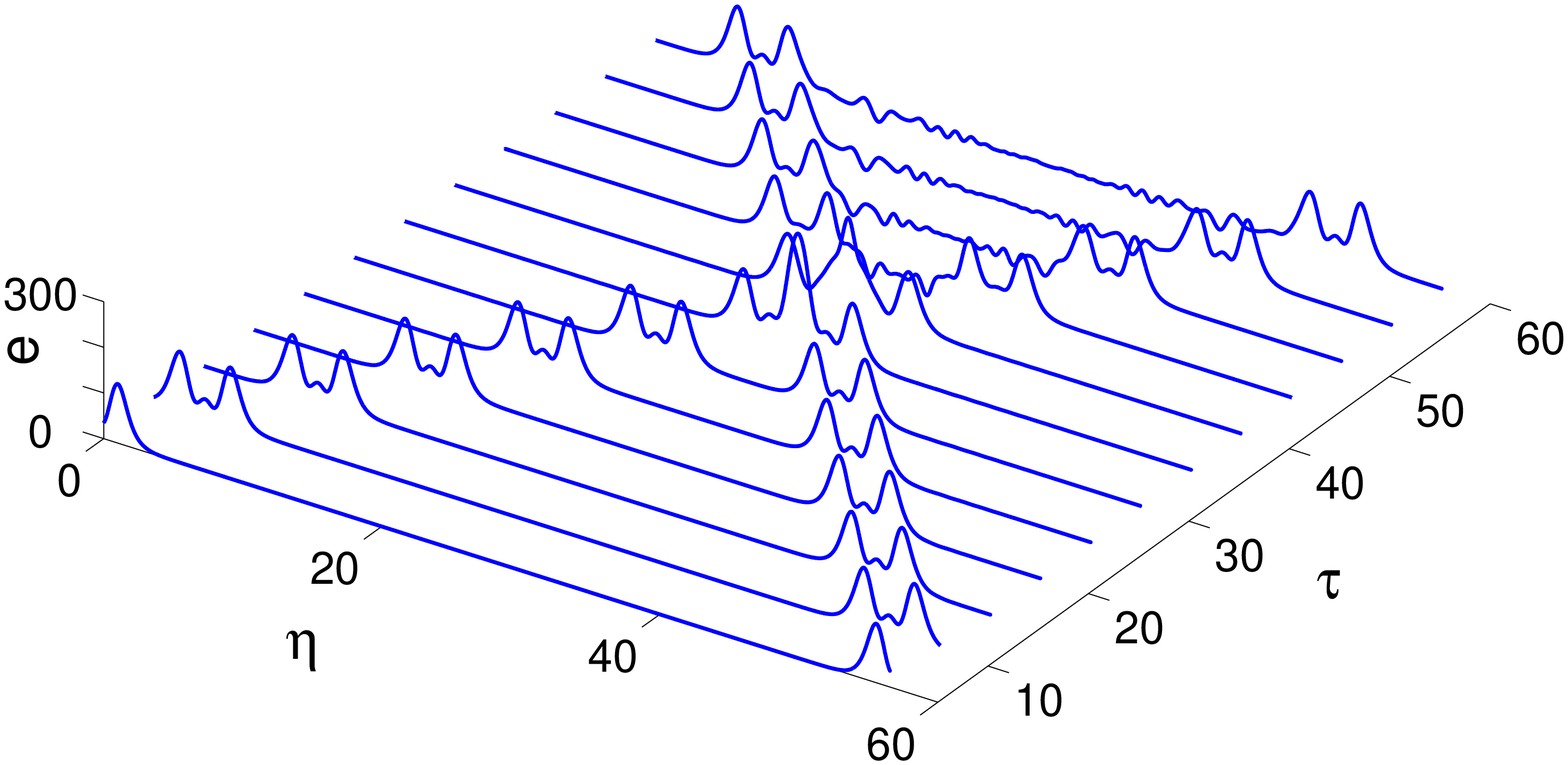}
\includegraphics [width=0.49\textwidth]{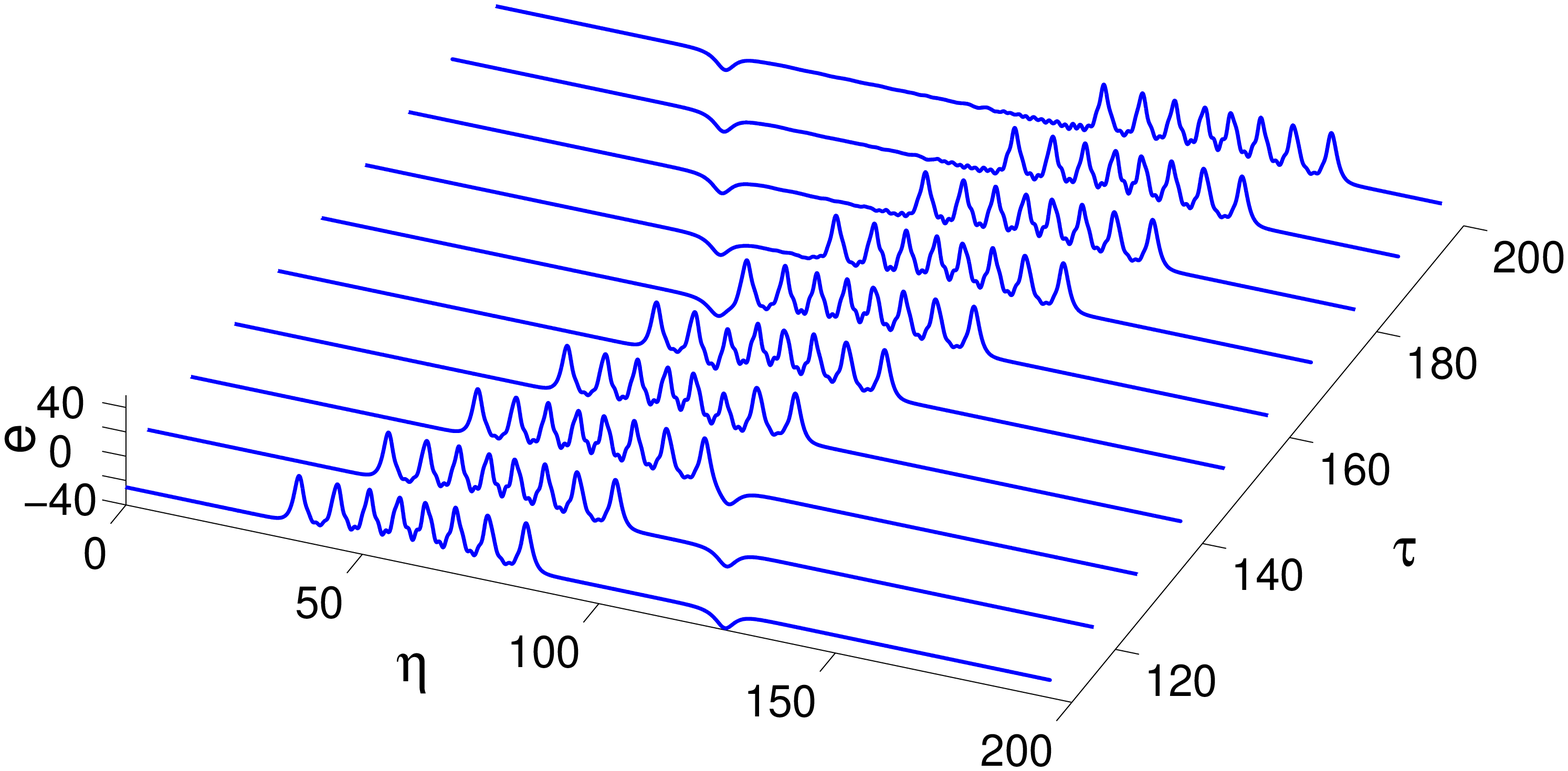}\caption{Solitary wave collisions:
two-hump solitons with $V=0.9$ and $V=-0.9$ (left); an eight-hump soliton with
$V=0.89$ and a phase-inverted soliton with $V=-0.75$ (right).}%
\label{Fig7}%
\end{figure}

Our numerical computations show that on a rather small interval $\frak{V}%
_{c}=[0.73,0.7642]$ at the low end of the spectrum, \textit{every} attempt of
computing a homoclinic orbit was successful (20 orbits per bin). These results
stay consistent with the refinement of the computational grid size $\delta V.$

All the solitary wave solutions in $\frak{V}_{c}$ are of the one-hump variety;
note, however that the single-hump solitons are not exclusively confined to
the lower end of $\frak{V}$. Elsewhere the spectrum density is very low, and
the solitons are mostly of a multi-hump kind. Somewhat arbitrarily, we define
a hump as a local maximum of the electric field $e,$ which is at least 50\% of
the global maximum.

Next we  studied the distribution of the different type of solitary wave
solutions
on the interval of velocities $[0.73,0.95]$. The figure
(Fig.~\ref{thehumpdistribution}) gives a very clear idea of the placement of
solitons according to the number of humps, which ranges from one to ten. Some
typical soliton profiles for four- and eight-hump solutions with $V=0.84322$
and $V=0.91416$ respectively are collected in Fig.~\ref{profiles:1-4}.

\begin{figure}[ht]
\centering
\subfigure[]{
\includegraphics[width=0.47\textwidth]{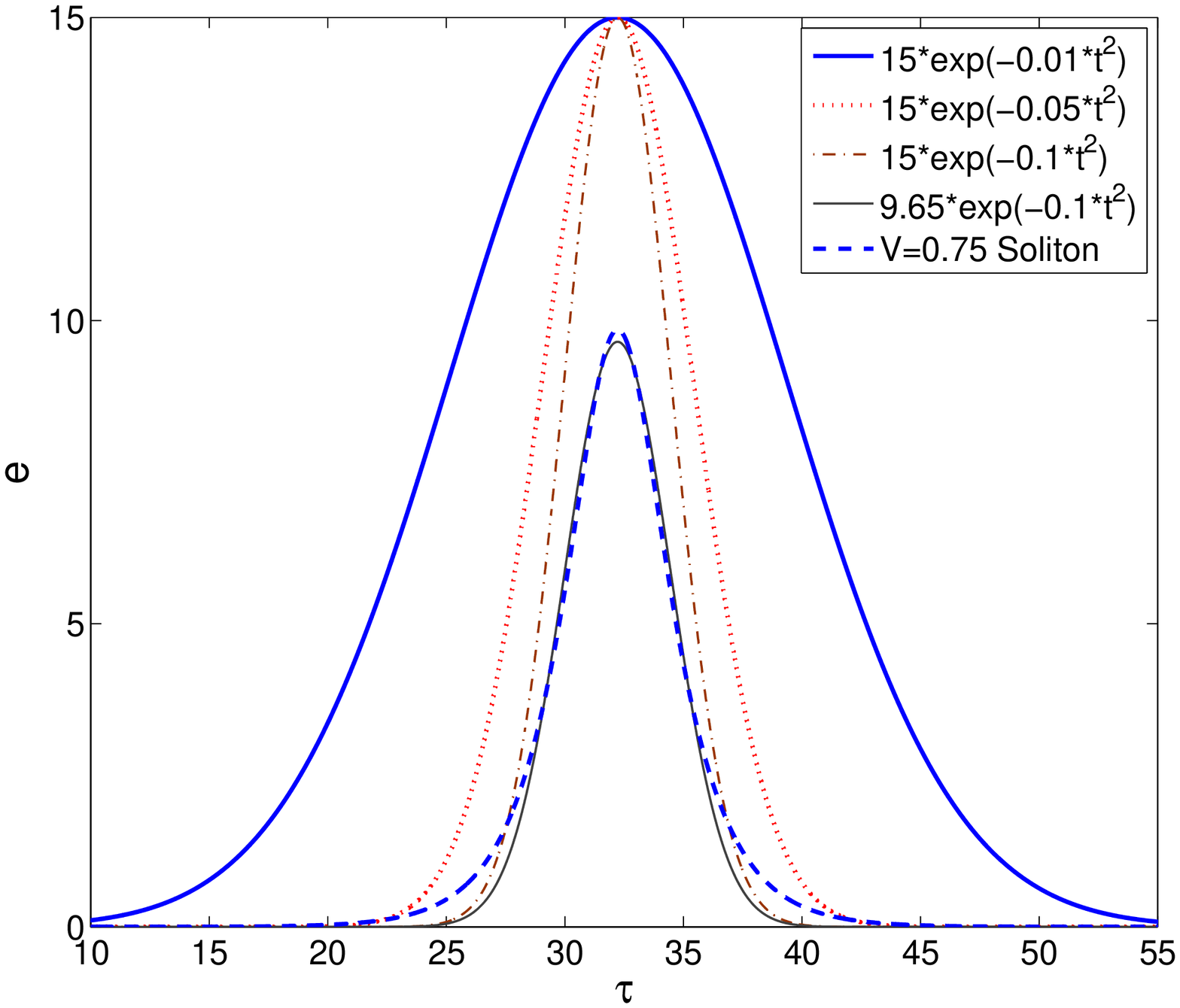}
\label{fig:energyic}
}
\subfigure[ ]{
\includegraphics[width=0.47\textwidth]{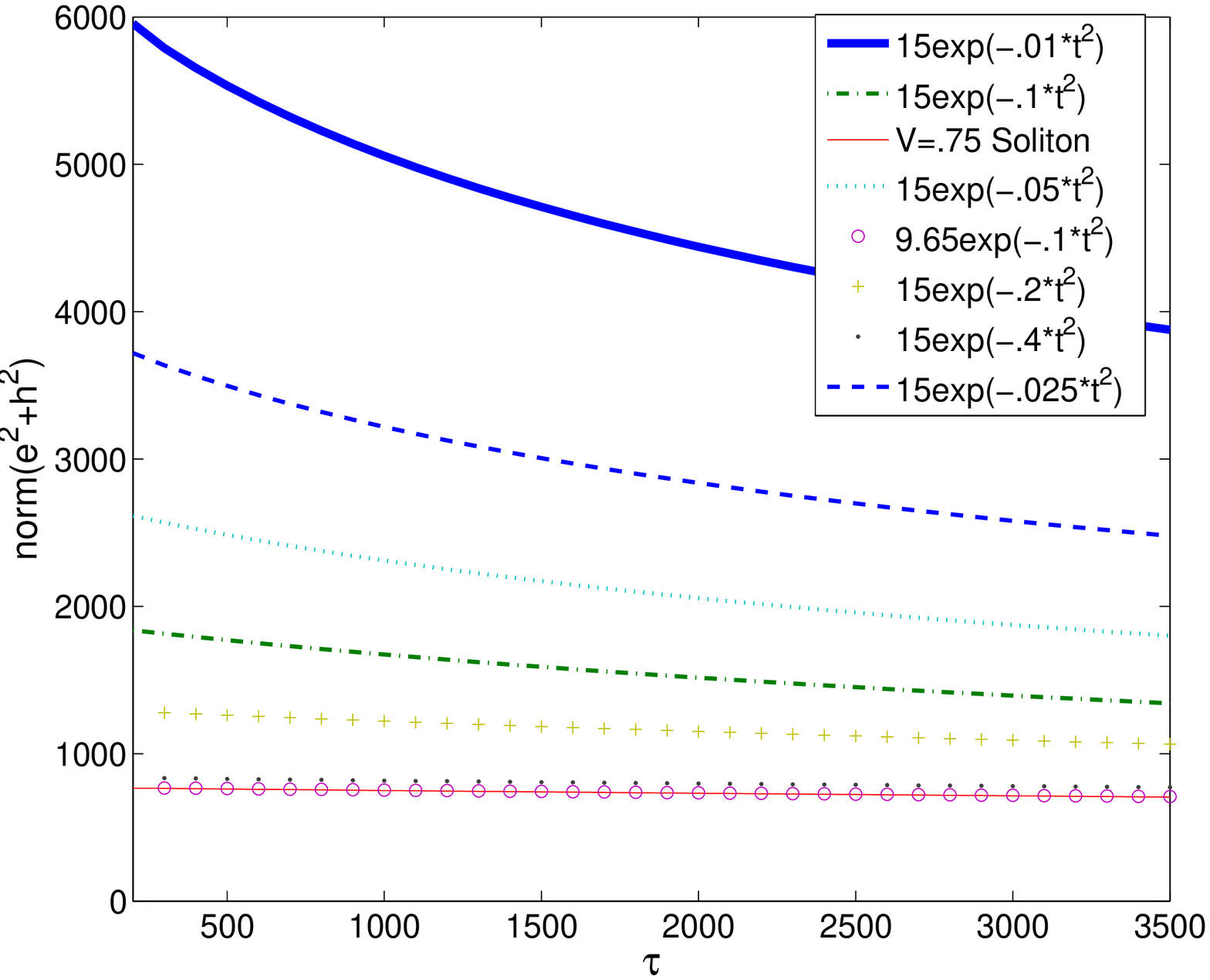}
\label{fig:energy}
}
\caption[]{a) Initial pulses for propagation study; b) Energy
dissipation for the given initial  pulses.  Larger Gaussian waves
quickly shed energy while breaking up into near-solitary waves. The near
solitary-waves slowly lose energy while converging to solitary waves.}
\end{figure}

Different types of solitary wave solutions have different energy values.
Because of the multi-hump nature of these solutions it is convenient to
introduce the energy of the electromagnetic field per one hump. We analyzed
dependence of the electromagnetic field energy $\mathcal{E}$ per one hump
versus velocity of solitary wave, see Fig.~\ref{Energy:per:single:hump}. Here
$\mathcal{E}$ is defined as
\[
\mathcal{E}=\frac{1}{2N}\int_{-\infty}^{\infty}\left[  e^{2}(t,x)+h^{2}%
(t,x)\right]  dt,
\]
where $N$ is the number of humps. As follows from this figure, in the log-log
coordinates the energy increase is very well approximated by a linear
function. The least square fit of the data from this figure shows that the
energy increases approximately as a polynomial of fifth degree in $V.$

\section{Formation, stability, and interaction of solitary waves: computer simulations}

In this section through direct numerical simulations we study evolution of
waves as well as wave interactions. We consider formation of solitary wave
solutions from arbitrary initial-boundary condition, stability of traveling
waves under small perturbations and stability under strong perturbations due
to wave collisions. In all numerical simulations of this section we use the
same values of physical parameters (\ref{params}) as in Sec.~\ref{num_waves}.

Numerically we solve the signaling problem for (\ref{dimensionless:system}).
In other words we give boundary conditions on either one or both ends of the
spatial interval $(0,L)$; as initial conditions we assign zero values for all
the variables, which corresponds to propagation in a quiescent medium. For
solving the initial-boundary value problem for the system
in~(\ref{dimensionless:system}) we devised a simple fractional step numerical
method.

Because the first two equations in (\ref{dimensionless:system}) are
hyperbolic \textit{PDEs} while the rest are \textit{ODEs} the choice of the
fractional steps is extremely natural: on the first half-step we propagate the
PDE part of the governing equations, and on the second half-step we march
according to the system of ODEs. The resulting ODE system is solved by using
the midpoint rule, while the PDEs are solved by the explicit McCormack method
\cite{laveque}. The midpoint rule and the McCormack method are both second
order accurate. To increase the accuracy of the fractional step method we
utilize the Strang split approximation \cite{strang}, which results in the
second order convergence of the final numerical scheme.

For many of the solitary-wave solutions discussed in Sec.~\ref{num_waves} we
ran direct numerical simulations on the model with these solitary waves as
input pulses. All the waves tested, even the ones of a rather intricate shape,
propagate with constant speed and without any shape distortion. See, for
example, Fig.~\ref{fig:EV6prop} where propagation of an eight-hump soliton is
depicted.

\begin{figure}[ht]
\begin{center}
\includegraphics[width=0.6\textwidth]{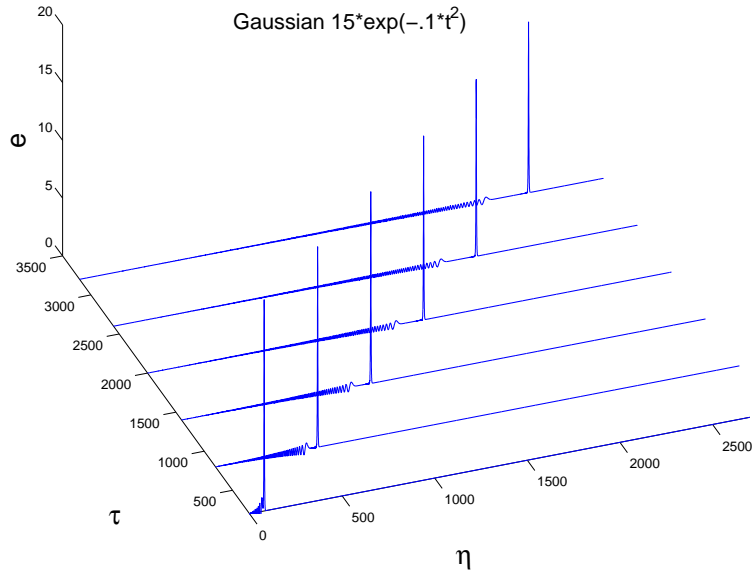}
\end{center}
\caption[]{Evolution of the $15\exp(-0.1\tau^{2})$ Gaussian.  A ``sharp'' Gaussian quickly
evolves into a near-solitary wave, leaving some disturbance in the wake. The
speed of the near-solitary wave is significantly higher than the speed of
propagation of the radiation; thus the wave quickly leaves the disturbance
behind }
\label{fig:goldgauss3d}
\end{figure}
This suggests that the solitary waves are (nonlinearly) stable with
respect to numerical perturbations. We remark that although because of the
scale of Fig.~\ref{fig:EV6prop}, the pulses appear rather singular, they are
in fact completely smooth and numerically resolved. The numerical resolution
of this computation is $8$ mesh intervals per unit length, which provides
about 60 mesh points per each hump of the traveling wave. Similarly, fine
computational meshes are employed in all the simulations below.

The issue of stability can be addressed analytically by studying the
linearization of the system of partial differential
equations~(\ref{dimensionless:system}) about arbitrary traveling wave
solutions and analyzing the corresponding linear evolution operator. Our
analysis showed that this operator is skew-Hermitian in ${L}_{2}$ with the
appropriate norm. Therefore the spectrum of the evolution operator is pure
imaginary and the traveling wave solutions are neutrally linearly stable (see
\cite{frenkel} for detail).

\begin{figure}[ht]
\begin{center}
\includegraphics[width=0.55\textwidth]{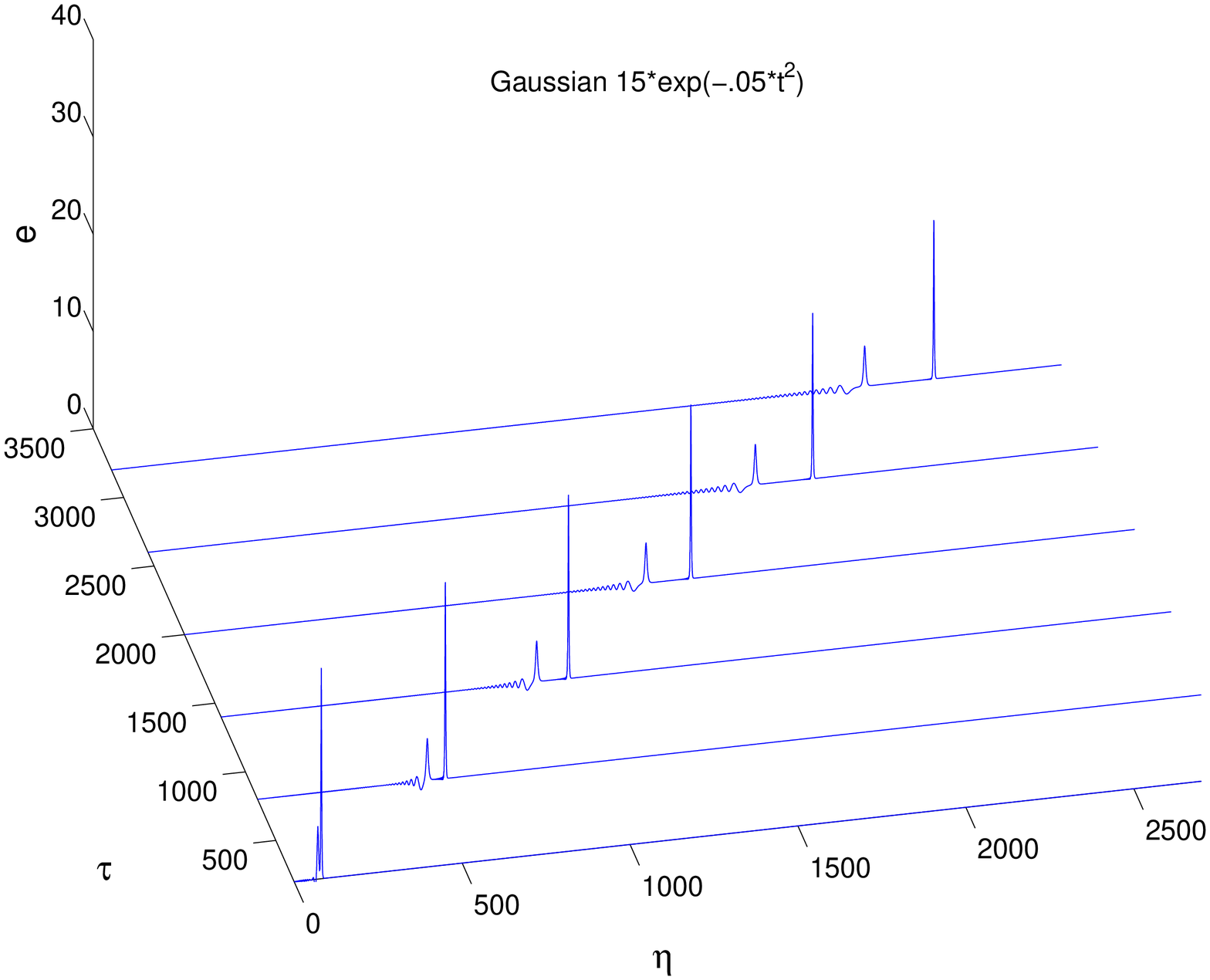}
\end{center}
\caption[ ]{Evolution of the $15\exp(-0.05\tau^{2})$ Gaussian. A medium size Gaussian evolves
into two waves. The velocity of the smaller wave is on the order of the
velocity of radiation. }
\label{fig:sharpgauss3d}
\end{figure}

To further elucidate the issue of stability we consider stability with respect to a finite-size perturbation in the initial wave shape. This situation is illustrated in Fig.~\ref{Fig6}. To the two-hump numerical soliton we add a rather substantial perturbation and employ the thus obtained functions as boundary data for the system of partial differential equations~(\ref{dimensionless:system}). As the result of evolution, the solution relaxes to the solitary wave shape followed by a low-amplitude ``continuous radiation''.

Stability with respect to strong perturbations due to collision of two traveling wave solutions is illustrated in Fig.~\ref{Fig7}. We take two solitary waves obtained by numerical solutions of ODEs and use these solutions as the boundary conditions for the PDEs. The left part of the Fig.~\ref{Fig7} shows collision of two-hump solitary wave solutions. The right part of the figure shows collision of eight-hump and one-hump solitary waves. In both cases collision of solitary waves leads to formation of the steady state solutions. The collisions are followed by emission of a small amplitude continuous radiation and a residual phase shift.

A soliton nature of solutions of (\ref{dimensionless:system}) is further confirmed by the set of numerical simulations we present next. We consider propagation of solutions with the pulses in Fig.~\ref{fig:energyic} given as a series of boundary conditions at the $x=0$ boundary. The soliton of velocity $V=0.75$ (see Fig.~\ref{Fig2}) propagates in a stable fashion, while its least-squares approximation by a Gaussian $9.65\exp{-0.1t^{2}}$ approaches the soliton shape after shedding a small amount of residual continuous radiation. These time evolutions are not included for space saving (the energy dissipation curves for these cases show conservation of electromagnetic energy, see Fig.~\ref{fig:energy}).

\begin{figure}{r}[ht]
\begin{center}
\includegraphics[width=0.55\textwidth]{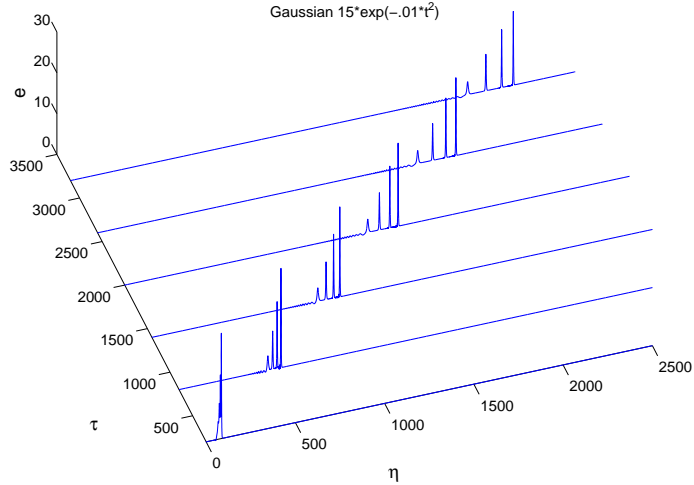}
\end{center}
\caption[]{Evolution of the $15\exp(-0.01\tau^{2})$
Gaussian.  Large Gaussian quickly breaks up into four near-solitary waves, leaving some disturbance in the wake. The waves become more separated over the time, since the near solitary waves with higher amplitude have higher velocities. }
\label{fig:widegauss3d}
\end{figure}

In the next three figures (Fig.~\ref{fig:goldgauss3d} -\ref{fig:widegauss3d}) we present evolutions of the larger Gaussian pulses from Fig.~\ref{fig:energyic}. Evolution of the sharpest Gaussian ($\sigma=\sqrt {5})$ is displayed in Fig.~\ref{fig:goldgauss3d}. Very fast the solution forms a solitary wave that moves with constant velocity with no shape change. It is followed by low magnitude oscillations whose leading edge also moves with constant speed. During the evolution, the oscillatory part disperses more and more. This part of the solutions appears to be of a nonlinear nature; it will be studied separately. We note that although because of the scale of the figure, the pulse appears very sharp, it is in fact completely smooth with ``width'' about $20$ and about $150$ computational mesh points within the pulse.

The evolution of a wider Gaussian, $\sigma=\sqrt{10}$ (see Fig.~\ref{fig:sharpgauss3d}) is similar with a very interesting distinction. Now the leading soliton is trailed by a slower low amplitude soliton. The latter is followed by low amplitude oscillations that again lag behind and disperse. The waves become more separated over time because the solitary waves with higher amplitude have higher velocities. Finally, the widest Gaussian, $\sigma=5\sqrt{2},$ develops into a train of four solitons, see Fig.~\ref{fig:widegauss3d}.

To characterize the energy exchange between the propagating pulse and the medium, in Fig.~\ref{fig:energy} we present plots of the total electromagnetic energy as a function of time for all the input profiles from Fig.~\ref{fig:energyic}. For the soliton solution there is a dynamic equilibrium between the energy stored in the medium and the electromagnetic energy of the pulse. In case of the input impulse being not a soliton, the balance between the medium and the pulse is violated, which leads to the dissipation of the electromagnetic energy into the medium.

\section{Concluding Remarks}

In this paper we considered propagation of extremely short pulses in a nonlinear medium, which is characterized by both electric and magnetic resonance responses. Interaction of the electromagnetic field with the medium was described in the framework of the Maxwell-Duffing model. In particular we employed  the classical Maxwell-Lorenz model for describing the magnetic resonance,~\cite{ZH01}. For describing the interaction of the electric field component with the medium we used a generalized Maxwell-Lorentz model which takes into account cubic anharmonism of the polarization response (i.e., the Maxwell-Duffing system). Our findings demonstrate that the model supports a wide array of traveling-wave solutions. We investigated the structure and properties of these solutions through a combination of analysis and numerical modeling. We determined that the family of traveling-wave solutions is parameterized by one parameter, which is the velocity of a steady wave solution, normalized by the speed of light in vacuum. The spectrum $\frak{V}$ contains both an interval of a continuous spectrum and a discrete subset of parameter values for which a traveling-wave solution exists. Computer modeling demonstrated a multi-hump structure of these solutions. Their multi-hump nature suggests to characterize solitary wave solutions by a number of humps (types). All types are determined by not overlapping sets of velocities.

Direct numerical simulations showed that solitary-wave solutions are dynamically stable. This dynamical stability is consistent with the analysis of the system linearized about solitary wave solutions~\cite{frenkel}. Stability of these solutions with respect to strong perturbations was studied
by means of solitary wave collisions. Computer simulations indicated nearly elastic nature of scattering followed by a residual excessive radiation and a phase shift. In addition to traveling-wave solutions, numerical simulations demonstrated presence of another type of nonlinear oscillatory solutions with extended tail.

\section*{Acknowledgment}

Frenkel's work was partially supported by the NSF EMSW21-RTG Grant No.~DMS-0636358. Part of this work is based on his Ph.D.~thesis \cite{frenkel}. This work was partially supported by NSF (grant DMS-0509589), ARO-MURI award 50342-PH-MUR, the State of Arizona (Proposition 301), and by the Russian Foundation for Basic Research through grant 06-02-16406. Roytburd's work was partially supported by the National Science Foundation, while working at the Foundation. Part of his work was performed during a sabbatical leave at the Lawrence Berkeley National Laboratory. The authors would like to thank M. Stepanov for the enlightening discussions and for the valuable help in preparation of this manuscript.

\end{document}